\documentclass[aps, prd, 11pt, showpacs, floatfix, superscriptaddress, twocolumn, nofootinbib, reprint, preprintnumbers, longbibliography]{revtex4-2}

\usepackage{lipsum, multirow, microtype, amsmath, amssymb, newfloat, bm, color}
\usepackage{graphicx}
\usepackage{natbib, hyperref}
\usepackage{hypcap, mathrsfs, placeins, etoolbox}
\usepackage{dcolumn}
\usepackage[dvipsnames]{xcolor}

\usepackage[T1]{fontenc}
\usepackage[scaled=0.92]{helvet}
\usepackage{inconsolata}
\usepackage{soul}

\usepackage{enumitem}
\setlist[itemize]{leftmargin=18pt, labelsep=0.45em, itemsep=0.065em}
\hypersetup{
  colorlinks=true,
  citecolor=red,
  linkcolor=magenta,
  urlcolor=NavyBlue,
  allbordercolors={0 0 0},
  pdfborderstyle={/S/U/W 1}
}

\newcommand{\norm}[1]{\ensuremath{\lVert#1\rVert}}

\def \Hz	    {\mathrm{Hz}}
\def \msun	    {{M}_\odot}
\def \flow	    {f_{\mathrm{low}}}
\def \d         {\boldsymbol{d}}
\def \h         {\boldsymbol{h}}
\def \noise     {\boldsymbol{n}}
\def \z         {z}
\def \param     {\vec \Lambda}
\def \intparam  {\vec \lambda}

\newcommand{\IITGn}{Indian Institute of Technology Gandhinagar, Gujarat 382355, India.\vspace*{4pt}}
\newcommand{\NIKHEF}{Nikhef, Science Park 105, 1098 XG Amsterdam, The Netherlands.}
\newcommand{\UU}{Institute for Gravitational and Subatomic Physics (GRASP), \mbox{Utrecht University}, Princetonplein 1, 3584 CC Utrecht, The Netherlands.}

\begin{document}
\title{Fast likelihood evaluation using meshfree approximations \mbox{for reconstructing compact binary sources}
}

\author{\sc{Lalit Pathak} 
} 
\email{lalit.pathak@iitgn.ac.in} 
\affiliation{\IITGn}
\author{\sc{Amit Reza}
} 
\email{areza@nikhef.nl} 
\affiliation{\NIKHEF} \affiliation{\UU}
\author{\sc{Anand S. Sengupta}
\vspace*{5pt}} 
\email{asengupta@iitgn.ac.in} 
\affiliation{\IITGn}
\begin{abstract}
Several rapid parameter estimation methods have recently been advanced to deal with the computational challenges of the problem of Bayesian inference of the properties of compact binary sources detected in the upcoming science runs of the terrestrial network of gravitational wave detectors. Some of these methods are well-optimized to reconstruct gravitational wave signals in nearly real-time necessary for multi-messenger astronomy. In this context, this work presents a new, computationally efficient algorithm for fast evaluation of the likelihood function using a combination of numerical linear algebra and mesh-free interpolation methods. 
The proposed method can rapidly evaluate the likelihood function at any arbitrary point of the sample space at a negligible loss of accuracy and is an alternative to the grid-based parameter estimation schemes. 
We obtain posterior samples over model parameters for a canonical binary neutron star system by interfacing our fast likelihood evaluation method with the nested sampling algorithm. The marginalized posterior distributions obtained from these samples are statistically identical to those obtained by brute force calculations. We find that such Bayesian posteriors can be determined within a few minutes of detecting such transient compact binary sources, thereby improving the chances of their prompt follow-up observations with telescopes at different wavelengths.
It may be possible to apply the blueprint of the meshfree technique presented in this study to Bayesian inference problems in other domains.
\end{abstract}

\pacs{}
\maketitle 

\section{Introduction}
\label{sec:intro}

The detection of gravitational waves (GW) from the GW170817~\cite{abbott2017gravitational} binary neutron star (BNS) system, followed by the prompt multi-wavelength (gamma-rays to radio) observation of its electromagnetic~(EM) counterpart, has led to several fundamental discoveries; and is hailed as a significant breakthrough in astronomy. These discoveries include the validation of long-held hypotheses that BNS mergers are ideal sites for r-process nucleosynthesis and produce short gamma-ray bursts, the first GW-based constraints on the equation of state of nuclear matter in such stars, and the measurement of Hubble constant independent of the cosmic distance ladder. 

The inevitable improvement of the detectors' sensitivity in future observation runs is likely to have a two-fold impact on the prospects of multi-messenger observations: 
{\emph{firstly}}, the increased bandwidth of the detectors (especially improved sensitivity at low frequencies) will result in a tremendous increase in the computational cost of Bayesian inference of source parameters, including sky localization essential for prompt observation of EM counterparts. Although the {\texttt{BAYESTAR}}~\cite{singer2016rapid} algorithm could be used to produce rapid sky maps, it has been recently shown~\cite{Finstad_2020} that coherent parameter estimation (PE) can localize the sources better by an average reduction of $14 \, \text{deg}^{2}$ in the uncertainty, underlining the importance of developing fast PE algorithms.
{\emph{Second}}, the reach of the terrestrial network of GW detectors will extend out to several Gpc to the effect that one would have far too many detections of BNS/NSBH signals to contend with whilst generating prompt sky-location maps~\cite{abbott2020prospects}; so much so that 
one may have to prioritize the GW sources for EM follow-up based on prospects of new science from a rapid estimation of their mass and spin components as shown by Margalit $\&$ Metzger~\cite{Margalit_2019}, thereby helping EM observatories to use resources optimally. 
Several fast PE algorithms have been developed recently, such as the coherent multi-detector extension of the relative binning/heterodyne method by Finstaad and Brown (2020)~\cite{Finstad_2020}, which produces the posterior within twenty minutes for BNS systems with 32 CPU cores. Well-trained machine learning PE methods~\cite{Dax2021, gabbard2022bayesian, Dax2023} can significantly reduce the runtimes and produce the posteriors in nearly real-time.
In the past, algorithms for accelerated parameter estimation have mainly focussed on speeding up the overlap integral. These include reduced-order models (ROMs)~\cite{canizares2015accelerated, Qi_2021, Soichiro_2020}, machine-learning aided ROMs~\cite{chua2020learning}, Gaussian process regression based interpolation~\cite{https://doi.org/10.48550/arxiv.1805.10457} and relative binning~\cite{cornish2021heterodyned, Venumadhav2018, Finstad_2020} algorithms. Our approach takes inspiration from the grid-based likelihood interpolation method~\cite{smith2014rapidly} based on orthonormal Chebyshev polynomials. The grid-based techniques have a drawback in that the number of interpolation nodes grows exponentially with the dimensionality of the parameter space.

In this work, we propose a new and alternative approach to grid-based likelihood interpolation method ~\cite{smith2014rapidly}, a computationally efficient method for evaluating the likelihood function (a key ingredient in Bayesian inference) using meshfree interpolation methods with dimension reduction techniques. We directly interpolate the likelihood function over the parameter space, bypassing the generation of templates and brute-force computation of the overlap integral altogether.  
Our scheme can quickly approximate the log-likelihood function with high accuracy and produce statistically indistinguishable posteriors over source parameters. 
Further, both the {\texttt{GstLAL}} search framework~\cite{GstLAL_2010} and the meshfree method use the idea of dimension reduction using SVD~\cite{golub2013matrix}, it may be prudent to incorporate this method with the low-latency {\texttt{GstLAL}} search pipeline for rapid, automated follow-ups of the detected events.

\section{Bayesian inference}
\label{sec:bayesianinference}

Given data ${\d = \h (\param_{\text{true}}) + \noise}$ recorded at a detector containing an astrophysical GW signal $\h (\param_{\text{true}})$ embedded in additive Gaussian noise $\noise$, one is interested in solving the inverse problem to estimate the source parameters. Bayesian inference is a stochastic inversion method where the posterior probability density $p(\param \mid \d)$ over the source parameters is related to the likelihood function ${\mathscr{L}(\d \mid \param)}$ of observing the data through the Bayes' theorem: 
\begin{equation}
p(\param \mid \d) = \frac{\mathscr{L}(\d \mid \param) \, p(\param)}{p(\d)} \, ,
\label{Eq:bayestheorem}
\end{equation}
where $p(\param)$ is the prior distribution over the model parameters ${\param \equiv \{ \intparam^{\text{ext}}, \intparam \}}$. 
In our notation, $\intparam$ denotes the intrinsic parameters such as component masses and spins. The set of extrinsic parameters is denoted by $\intparam^{\text{ext}}$.
We are particularly interested in estimating the extrinsic parameter $t_{c}$ denoting the fiducial time of coalescence of the two masses. $t_{c}$  will be mentioned explicitly wherever required, as it is treated in a special way in our analysis.

The forward generative frequency-domain restricted waveform model for non-precessing compact binaries can be expressed  as ${\h (\param) = {\mathcal{A}}\, h_+(f_{k}; \vec\lambda)}$, where the complex amplitude ${\mathcal{A}}$ depends only on the extrinsic parameters and $h_+(f_{k}; \vec\lambda)$ is the `+' polarization of the signal that depends only on the intrinsic parameters~\cite{Foreman_Mackey_2013}. Here ${\{f_{k}\}_{k = 0}^{N_{s}/2}}$ defines positive Fourier frequencies, and $N_s$\footnote{${N_{s} = \text{signal duration} \times \text{sampling frequency}}$} is the number of sample points.
A GW signal, observed by an interferometric detector, can be considered as a linear combination of the two polarizations weighted by the antenna pattern function. The $h_{+}(f_{k}; \vec\lambda)$ polarization is related to the $h_{\times}(f_{k}; \vec\lambda)$ polarization for non-precessing GW signal as: ${h_{+}(f_{k}; \vec\lambda) \propto i h_{\times}(f_{k}; \vec\lambda)}$~\cite{findchirp_2012}. This allows us to write the detector response in terms of any one of the polarizations alone (we have chosen the $h_{+}(f_{k}; \vec\lambda)$ polarization). 
Using this model, the posterior ${p(\param \mid \d)}$ can be directly evaluated at every point in $\param$ using Eq.~(\ref{Eq:bayestheorem}). However, in view of the high-dimensionality of $\param$, it is more efficient to sample the posterior using stochastic sampling algorithms such as Nested-Sampling~\cite{skilling2006nested}, or Markov Chain Monte Carlo (MCMC)~\cite{Foreman_Mackey_2013}. 
From Eq.~(\ref{Eq:bayestheorem}), it is evident that for a quick estimation of the posterior distribution, it is imperative to rapidly evaluate the likelihood function.

We work with the phase-marginalized log-likelihood function~\cite{thrane_2019}:
\begin{equation}
\begin{split}
\ln \mathscr{L} (\param, t_{c})	= \ln I_{0} \left [ |{\mathcal{A}}| \, z(\vec\lambda, tc) \right ] 
								- \frac{1}{2} \norm{\h(\param)}_2^2 
\end{split}
\label{Eq:log-likelihood}
\end{equation}
where $I_0(\cdot )$ is the 0-th order modified Bessel function of the first kind, 
and $\z(\vec\lambda, t_{c})$ is the frequency-domain overlap-integral: 
\begin{equation}
\z(\vec\lambda, t_c) = 4 \, \Delta f \, \left | \sum_{k = 0}^{N_{s}/2} {
\frac{{d}^{*}(f_{k}) \, {h}_+(f_{k}, \vec\lambda)}{S_{h}(f_{k})} \, e^{-2 \pi i f_{k} t_{c}}}  \right |,  
\label{Eq:overlap}
\end{equation}
inversely by $S_{h}(f_{k})$, the detector's one-sided noise power spectral density (PSD).
The data and template vectors are sampled at discrete frequencies $\{f_{k}\}_{k = 0}^{N_{s}/2}$. 

The complexity of evaluating the overlap integral scales directly with the number of data samples, which in turn, scales  with the seismic cut-off frequency (approximately) as ${N_s \sim \flow^{-8/3}}$.  
As we progress from the O4 observational run (${\flow = 20\,\Hz}$) to O5 at design sensitivity (${\flow = 10\,\Hz}$), evaluating ${p(\param \mid \d)}$ is likely to take  at least $\times 6.3$ longer. In addition, additional costs will be incurred in constructing longer templates at the proposal points. Therefore, the likelihood calculation can be expensive. However, our method is immune to this issue as our scheme directly approximates the likelihoods at different sample points.
In this work, we have used non-orthonormal radial basis functions (RBF) (Gaussian kernels) centered at interpolation nodes that can be randomly scattered over the volume of the intrinsic parameter space.  
In this manner, we have effective control of their number in higher dimensional parameter spaces. 

\section{Meshfree likelihood interpolation}
\label{sec:meshfree_likelihood_interpolation}

The computational cost of Bayesian inference comprises of two parts: the first is incurred in waveform generation followed by likelihood evaluation at a point proposed by the sampler. Typically, a sampler proposes a large number (${\sim 10^6 - 10^7}$) of points to adequately capture the posterior distribution - which makes this part computationally expensive. The other part of the total computational cost can be attributed to the overheads of the sampling method itself. Since the latter cost depends on the efficiency of the sampling algorithm used (and its software implementation) and is significantly less in comparison to the overall cost of PE, we shall ignore it in our discussions. 

We assume that the parameter estimation is ``seeded'' by the most significant trigger $\param^\ast$ from an upstream detection pipeline~\cite {messick2017analysis, usman2016pycbc}. The sampling algorithm draws new proposals from a sample space which is taken to be a moderate-sized ``hyper-rectangle'' in $\param$, centered around the most significant search trigger. 

From Eq.~\eqref{Eq:log-likelihood} and~\eqref{Eq:overlap}, it is evident that we need to interpolate two pieces: $\z(\intparam, t_{c})$ and $\norm{h_+(\intparam)}_2^2$; and combine them with the amplitude ${\mathcal{A}}$ to calculate the log-likelihood ratio at a given `query' point $\intparam^q$ and a particularly given value of $t_c$.  For this purpose, a set of $n$ unique interpolation nodes $\intparam^\alpha, \ \alpha = 1, \cdots, n$ are randomly chosen from a uniform distribution over the sample space. 

The template norm $\norm{h_+(\intparam)}_2^2$ is a smoothly varying scalar field over $\intparam$. We can interpolate its value at a query point $\intparam^q$ by first evaluating the values explicitly at the interpolation nodes $\intparam^\alpha$, and then expressing $\norm{h_+(\intparam^q)}_2^2$ at an arbitrary point as a linear combination of Gaussian RBF kernels centred at these nodes. The unknown coefficients of this linear combination can be uniquely found by enforcing the interpolation criteria as explained in the next section.

On the other hand, as the overlap integral has to be evaluated at an arbitrary point $(\intparam^q, t_{c})$, it will turn out to be more convenient to interpolate it as a vector. In this case, a set of overlap-integral vectors $\vec{z}_\alpha$ are first constructed at the interpolation nodes, sampled on a uniform grid over $t_c$. The vectors $\vec{z}_\alpha$ have elements $z_\alpha[k] \equiv z(\intparam^\alpha, k \, \Delta t)$,  where $\Delta t$ is the sampling interval, $k$ is an integer $\in {\text{int}} ([t_c^\ast \pm \tau]/\Delta t)$, $t_c^\ast$ represents the `reference' coalescence time as triggered by the search pipeline and $2\tau$ is the dimension of the sample-space (hyper-rectangle in $\param$) along the $t_{c}$ direction.

Since the bulk of the support for the posterior distribution comes from near the peak of these time series, we choose its samples that are centred around the triggered value. From Eq.~\eqref{Eq:overlap} it is clear that $\vec{z}_\alpha$'s can be constructed efficiently using FFT correlations. Once the set of vectors $\{ \vec{z}_\alpha \}$ is available, they can be projected over a suitable set of basis vectors $\{ \vec u_\mu \}$ with linear coefficients $\{C_\mu(\intparam^\alpha)\}$.  
As each of the coefficients is smoothly varying scalar fields over $\intparam$ (sampled at the interpolation nodes), we can use meshfree  methods to interpolate their values at an arbitrary query point. 
The meshfree scheme can be divided into two stages: (i) a preparatory, start-up stage where we explicitly determine the RBF interpolating functions (interpolants) from the pre-computed likelihood values at the interpolation nodes and (ii) an online stage where these interpolants are evaluated on the fly to `predict' the likelihood values at arbitrary query points in the sample space.

\subsection{Start-up stage}
\label{subsec:startup}

In this stage, the nodes are randomly sprayed over the sample space, and the meshfree interpolants are constructed.

\begin{itemize} 
\item[1.] {SVD \emph{basis}}:
We are interested in finding a suitable set of basis vectors that span the space of $n$ input overlap vectors $\{\vec{z}_\alpha\}$. This is conveniently performed by stacking these vectors row-wise and performing a singular-value decomposition (SVD) of the resultant matrix: 
\begin{equation}
	\vec \z_\alpha  =
				\sum_{\mu = 1}^{n} C^{\alpha}_\mu  \, \vec {u}_{\mu} \, ,
\label{Eq:SVD}
\end{equation}
where $C^{\alpha}_{\mu}$ are the coefficients for the set of orthonormal basis vectors $\vec{u}_\mu$ in decreasing order of their relative importance as determined from the spectrum of singular values. 
A strong correlation between the $\vec{z}_\alpha$'s implies that the overlap vectors lie in the span of 
the top-$\ell$ basis vectors ($\ell \ll n$).
A vector $\vec{z}_q$ at a query point in the sample space can also be spanned by the same set of basis vectors. 

Note that for a fixed index $\mu$, the coefficients $C^{\alpha}_\mu$ represent a surface whose values are known only at the input nodes. Along with Eq.~\eqref{Eq:SVD}, this implies that the interpolation of the inner-product vector at an arbitrary query point essentially boils down to interpolating the value of the coefficients $C^q_{\mu} \equiv C_\mu(\intparam^q)$. 

\item[2.] {\emph{Creating meshfree interpolants}}: 
In this step, we create and explicitly determine the `meshfree' interpolants for each of the coefficient surfaces (that appear in Eq.~\eqref{Eq:SVD}) independently. The interpolant for the coefficient corresponding to the $\mu^{\text{th}}$ basis vector can be expressed~\cite{doi:10.1142/6437} as a linear combination of RBF kernels centred on the scattered, distinct nodes $\intparam^\alpha$, augmented by monomials ranging up-to a specific order:
\begin{equation}
    C^{q}_{\mu} = \sum_{\alpha = 1}^{n} r_{\alpha} \, \phi(\norm{\intparam^q - \intparam^\alpha}_2) \, 
        									+ \,\sum_{j = 1}^{M} b_{j} \, f_j(\intparam^q) \, ,
\label{Eq:rbfcoeffs}
\end{equation} 
where $\phi$ is the Gaussian kernel centred on $\intparam^{\alpha}  \in \mathbb{R}^d$, and $\{f_j\}$'s are monomials that span the space of polynomials of some preset target degree $\nu$ in $d$ variables.  Also, ${\boldsymbol{r} = [r_{1}, r_{2}, \ldots, r_{n}]^T}$ and ${\boldsymbol{b} = [b_{1}, b_{2}, \ldots, b_{M}]^T}$ are the set of $(n+M)$ coefficients that need to be uniquely determined to determine the interpolant.  

Since $C^{q}_{\mu}$ are known at the interpolation nodes, it allows us to enforce $n$ interpolation conditions. $M$ additional conditions $\sum_{k=1}^n r_k \,f_j(\intparam^k) = 0, \, j=1, \ldots, M$ are added to ensure a unique solution. Together, these lead to a system of equations:
\begin{equation}
	\begin{bmatrix}
		\boldsymbol{K} & \boldsymbol{F} \\
		\boldsymbol{F}^{T} & \boldsymbol{O} 
	\end{bmatrix} \
		\begin{bmatrix}
		\boldsymbol{r} \\ 
		\boldsymbol{b}
	\end{bmatrix} \ 
	=  
	\begin{bmatrix}
		C^{\alpha}_{\mu} \\ 
		\boldsymbol{0} 
	\end{bmatrix}
\label{Eq:RBF-SLE}
\end{equation}
where the matrices $\boldsymbol{K}$ and $\boldsymbol{F}$ have components $K_{ij} = \phi(\norm{\intparam^{i}-\intparam^{j}}_2)$ and $F_{ij} = f_j(\intparam^{i})$ respectively; $\boldsymbol{O}_{M\times M}$ is a zero-matrix and $\boldsymbol{0}_{M\times 1}$ is a zero-vector. Eq.~\eqref{Eq:RBF-SLE} can be solved uniquely for the unknown coefficients $\boldsymbol{r}$ and $\boldsymbol{b}$, thus completely determining the meshfree interpolant in Eq.~\eqref{Eq:rbfcoeffs}. The solution for $\boldsymbol{r}$ and $\boldsymbol{b}$ can be shown to be unique if $\boldsymbol{F}$ has full column rank.

A minimum set of ${n = \binom{\nu + d}{\nu}}$ interpolation nodes are required to be uniformly distributed over the $d$-dimensional intrinsic parameter space to determine the coefficients uniquely. 
Euclidean distances between pairs of points seem to work well, possibly due to the small volume of the sample space in a typical PE analysis. Parameter-space metric-based distances could also be used.

A similar procedure is followed to create a separate meshfree interpolant for $\norm{h_+(\intparam^{q})}_2^2$.
 
\end{itemize}
Note that by construction, the meshfree interpolation eliminates both the factors that contribute to the high computational cost, namely, (a) generating the forward signal model and (b) explicitly calculating the overlap integral at every query point proposed by the sampling algorithm. 

\subsection{Online stage}
\label{subsec:online}

In this stage, the $(\ell+1)$ interpolants (prepared in the offline stage earlier) are evaluated on the fly at arbitrary query points $(\intparam^q, t_c)$ proposed by the sampling algorithm. 

The interpolant for the square of the template norm can be directly evaluated to get the interpolated value $\norm{h_+(\intparam^{q})}_2^2$. Similarly, the $\ell$ interpolants for SVD coefficients (Eq.~\eqref{Eq:rbfcoeffs}) are evaluated to get a set of interpolated coefficients, which are then combined with the corresponding top-$\ell$ basis vectors $\vec{u}_\mu$ (see Eq.~\eqref{Eq:SVD}) to obtain the interpolated overlap-integral $\vec{z}_q$. 

By construction, $\vec{z}_q$ is uniformly sampled over the interval $[t_c^\ast \pm \tau]$. As such, it is possible that the query $t_c$ does not coincide with the discrete-time samples of $\vec{z}_q$. In such a case, we use a one-dimensional cubic-spline interpolation to evaluate ${z}(\intparam^q, t_c)$ using a few `nearby' grid samples of $\vec z_q$ as input. This implies that $\vec z_q$ has to be reconstituted only at a few ($\sim 10$) consecutive sample points, which considerably accelerates the matrix-vector multiplication in Eq.~\eqref{Eq:SVD}.

Combining the interpolated values $z(\intparam^q, t_c)$ and $\norm{h_+(\intparam^q)}_2^2$ with the extrinsic-parameter dependent complex amplitude ${\mathcal{A}}$ (see Eq.~\eqref{Eq:log-likelihood}), we finally obtain the log-likelihood ratio $\ln \mathscr{L}$ at the arbitrary point $(\param^q, t_{c})$.

\section{Numerical experiments}
\label{sec:numerical_experiments}

To demonstrate the accuracy and speed of the meshfree method in reconstructing the source parameters, we prepared synthetic $360 \, \text{s}$ long data $\d$, sampled at $4096 \,\Hz$. For this, a simulated GW signal $\h$ from a canonical BNS system with component masses $m_{1, 2} = 1.4 \msun$ and mass-weighted effective dimensionless spin $\chi_{\text{eff}} = 0.05$ was injected (using the {\texttt{IMRPhenomD}}~\cite{khan2016frequency} signal model) in colored Gaussian noise using the noise power spectral density model~\cite{aLIGO_ZDHP} of {\texttt{aLIGO}} detectors. 
The distance to the source was adjusted for a moderate matched-filtering SNR of $10$.
The seismic cut-off frequency was chosen to be $20 \, \Hz$ to mimic data from the upcoming O4 science run.

\begin{table}[hbt]
\def\arraystretch{1.47}
\begin{ruledtabular}
\begin{tabular}{ c|c|c|c|rd} 
& $ \mathcal{M}/M_{\odot}$ & $\eta$ & $\chi\strut_{\text{eff}}$ &  $\text{SNR}$\\ \hline
Injection   & $1.2187$ & $0.25$ & $0.05$ & $10.00$  \\ \hline
Standard    & $1.2187\;\strut^{1.2188}_{1.2185}$ & $0.2487\;\strut^{0.2498}_{0.2456}$ & $0.050\;\strut^{0.051}_{0.049}$ & $9.67$  \\ \hline
Meshfree    & $1.2187\;\strut^{1.2188}_{1.2185}$ & $0.2487\;\strut^{0.2498}_{0.2456}$ & $0.050\;\strut^{0.051}_{0.049}$ & $9.67$  \\
\end{tabular}
\caption{Reconstruction of a canonical BNS event.}
\label{tab:Inj_details}
\end{ruledtabular}
\end{table}

We performed Bayesian inference on this simulated data using both (a) {\emph{direct}} likelihood calculation used in PyCBC inference and (b) and by using the proposed meshfree likelihood interpolation scheme outlined in earlier sections. We used publicly available software~\cite{RBF_github} for radial basis functions and the {\texttt{Dynesty}}~\cite{speagle2020dynesty, sergey_koposov_2023_7600689} nested-sampling package for carrying out the Bayesian inference analysis.
We varied four intrinsic parameters (component masses and aligned-spin magnitudes) and two extrinsic parameters (luminosity distance and coalescence time), keeping other parameters fixed.

For this exercise, we used $n = 800$ random input nodes over $\intparam$. The top $\ell = 120$ basis vectors and a polynomial order $\nu = 6$ with a corresponding nominal median relative error $\sim 10^{-5}$ across the sample space in approximating the log-likelihood function.
The accuracy trade-offs of likelihood reconstruction by varying basis size is shown in the Appendix~\ref{appendix:A}.

The meshfree parameter reconstruction was completed in $5.3$~min in comparison to $31.7$~h taken by the direct calculation. The likelihood function was evaluated $676$ times faster using the meshfree method. 

\begin{figure}[htb!]
\includegraphics[width=\columnwidth, clip=True]{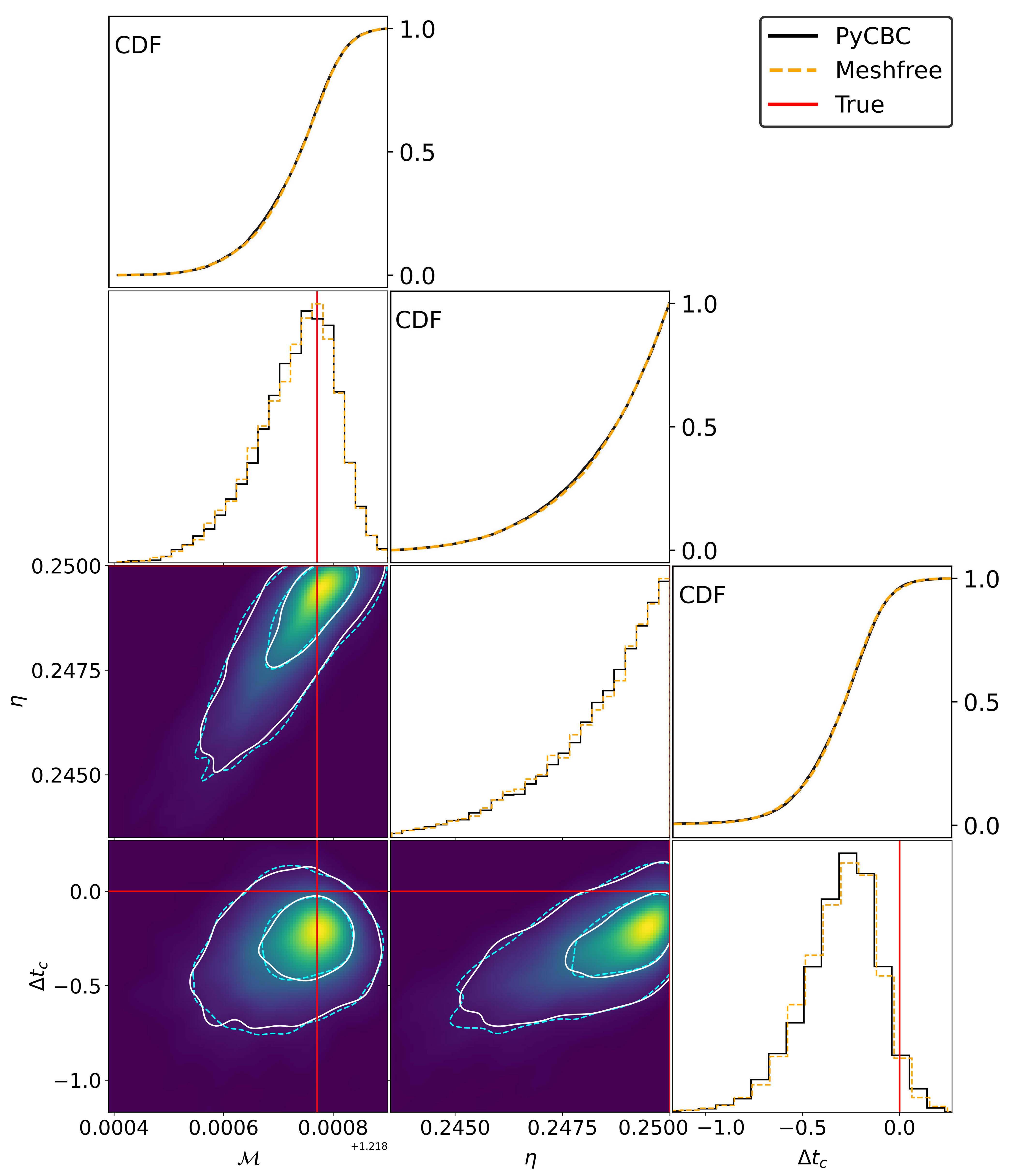}
\caption{\label{Fig:Corner-plot}
Marginalized PDF for chirp mass $\mathcal{M}$, symmetric mass ratio $\eta$, and $\Delta \text{t}_{c}$ parameters of a simulated BNS event at a seismic cutoff of $20\, \Hz$. The injection parameters are shown as red lines. The $50\%$ and $90\%$ contours for meshfree method (dashed cyan trace) and PyCBC (solid white trace) are also shown. The plot-overlaid marginalized PDF obtained from the proposed meshfree method (dashed orange trace) and standard PyCBC inference (black line) are virtually indistinguishable. 
}
\end{figure}

Some of the estimated parameters have been compared in Table~\ref{tab:Inj_details}, which show identical values obtained by both methods. 
The marginalized PDF over three parameters $\mathcal{M}$, $\eta$, and $\Delta t_{c}$ are shown in the corner plot Fig.~\ref{Fig:Corner-plot}. The figure also contains cumulative density (CDF) profiles of these  distributions obtained from both PyCBC inference and the meshfree method plotted together. 
While both the PDF and CDF profiles look virtually indistinguishable, we also calculate statistical measures of similarity~\cite{stats_book} using the PDFs obtained from the two methods for further validation: both the Kolmogorov-Smirnov statistic ($0.0130$) and the Bhattacharyya distance ($0.0006$) between the chirp-mass PDF profiles support the fact that the two distributions are nearly identical. We get  similar results for posterior distributions of other parameters.

This numerical example shows that the meshfree method can generate a statistically indistinguishable replica of the posterior distributions in a GW Bayesian inference problem at a small fraction of the total computational cost.

\section{Speed-up analysis}
\label{sec:speed-up}

We calculated the ratio of (average) time taken to compute the log-likelihood by these techniques at a fixed accuracy of reconstruction. Several simulated data sets were used in this study, generated by injecting signals having different parameters in colored Gaussian noise using the {\texttt{aLIGO}} noise model at a fixed matched-filtering SNR of $10$.  Seismic cutoff frequencies at $20\,\Hz$ ($10\,\Hz$) were considered to mimic data from upcoming O4 (O5) science runs. 

\begin{table}[hbt]
\def\arraystretch{1.075}
\begin{ruledtabular}
\begin{tabular}{p{0.25\columnwidth}|d|d|d|rd} 
Seismic cut-off 
	& \multicolumn{1}{c|}{$M /M_{\odot}$} 
	& \multicolumn{1}{c|}{$t_\text{mf}$ /ms} 
	& \multicolumn{1}{c|}{$t_\text{pycbc}$ /ms} 
	& \multicolumn{1}{c}{speed-up} \\ 
\hline
\multirow{3}{*}{$\flow = 10\,\Hz$}
	& 2.8 & 0.90 & 3604.76 & 4005.2 \\
	& 4.0 & 1.06 & 1405.84 & 1326.2 \\
	& 20.0 & 0.88 & 29.27 & 33.2 \\ 
\hline
\multirow{3}{*}{$\flow = 20\,\Hz$}
	& 2.8 & 0.67 & 452.93 & 676.0 \\
	& 4.0 & 0.81 & 207.35 & 256.0 \\ 
	& 20.0 & 0.63 & 3.32 & 5.3 
\end{tabular}
\caption{\label{tab:speed_up}
The first and second columns define the seismic cut-off frequency and total mass of the injected GW signal, respectively. The third and fourth columns show the median time (in ms) taken for a single evaluation of the log-likelihood function using the standard PyCBC method and our approach at a nominal relative error of $\mathcal{O}(10^{-5})$. The last column provides the relative speed-up of our method in comparison with the standard likelihood calculation. Median time (in ms) taken for a single evaluation of the log-likelihood function using standard PyCBC method and the meshfree approach at a nominal relative error of $\sim 10^{-5}$. 
}
\end{ruledtabular}
\end{table}

There is an obvious trade-off between accuracy and speed-up of the meshfree method, which are determined by the choice of $(n, \ell, \nu)$ parameters. Larger values can lead to more accurate  likelihood estimates, albeit at a higher computational cost and vice-versa. 
We used a heuristic combination to guarantee median (relative) errors $\lesssim 9 \times 10^{-5}$ in the estimated log-likelihood values across the entire sample space. 

\begin{figure}[t]
\includegraphics[width=\columnwidth, clip=True]{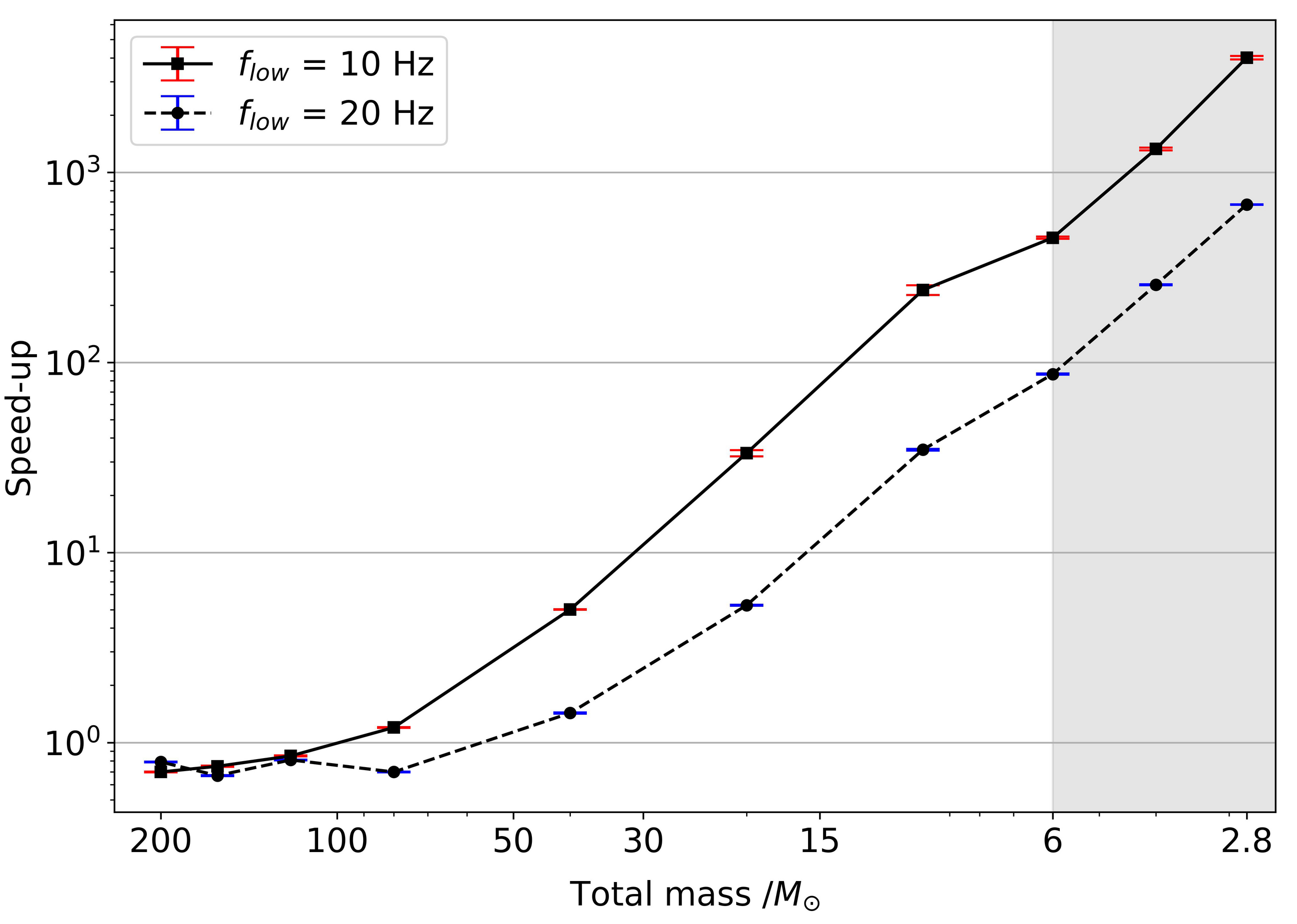}
\caption{\label{Fig:speed_up_plot}Speed-up factors (averaged over several thousand evaluations) in calculating the log-likelihood function with the meshfree method for different equal-mass ($q=1$) systems with (dimensionless) spin magnitude of $0.05 \, (0.2)$ for BNS (BBH) systems. The shaded region denotes BNS systems. The combination of $(n, \ell, \nu)$ parameters were tuned for a median error of $\sim 10^{-5}$.  
}
\end{figure}

The log-likelihood function was evaluated and timed for a large number of random points uniformly  distributed over the $(\intparam, t_{c})$ space. 
Table~\ref{tab:speed_up} summarizes the speed-ups corresponding to the two different seismic cutoff frequencies for three compact binary systems with equal component masses. It is further elucidated in Fig.~\ref{Fig:speed_up_plot}, where the speed-up comparison is drawn between a larger number of equal-mass binary systems covering a wider range of parameters. From Table~\ref{tab:speed_up} and Fig.~\ref{Fig:speed_up_plot}, it is clear that the likelihood computation can be sped up $\sim 4000$ times faster for canonical BNS systems at a nominal error of $\sim 10^{-5}$ using the meshfree method against standard likelihood implementation in PyCBC inference. However, this would not reflect the true benefit of our proposed method, as PyCBC inference is not optimized to calculate the fast posterior distribution for LIGO. 
Our method computes the posterior within a few minutes for BNS system. Thus, our proposed scheme has the potential to perform rapid reconstruction of source parameters in upcoming observation runs for {\texttt{aLIGO}} detectors similar to the other existing optimized methods.

The low-mass systems with a large number of in-band cycles would benefit most from the meshfree method.
In contrast to the standard method, the time taken by the meshfree method is relatively unaffected by the chirp time of the signals. All the tests were performed on a single-core AMD EPYC 7542 CPU@2.90GHz CPU. \\

%

\section{Conclusion and Outlook}
\label{sec:conclusion}

We have presented an alternative approach to the grid-based method~\cite{smith2014rapidly}, which is computationally efficient for accurately evaluating the log-likelihood function. 
Our meshfree method can be easily integrated into well-known sampling algorithms (e.g., MCMC, Nested Sampling) to substantially accelerate the Bayesian inference of source parameters of coalescing compact binary sources, triggering prompt observation of their EM counterparts in the future. 
Using synthetically generated data of a GW signal from a merging BNS system, we have demonstrated that the posterior distributions are statistically identical to those obtained using the standard PyCBC inference ~\cite{biwer2019pycbc}. 
For BNS systems, the likelihood function can be calculated $\sim 4000$ times faster at any point proposed by the sampling algorithm than the direct likelihood calculation implemented in PyCBC inference.
At this point, we want to remind the readers that we made the comparison of our scheme against PyCBC inference only to verify the robustness of our scheme. PyCBC inference is not used for the fast PE run by LVK. Therefore, for a fair speed-up comparison, we must compare our methods against those schemes ~\cite{Finstad_2020, Dax2021, gabbard2022bayesian, canizares2015accelerated, Qi_2021, Soichiro_2020, https://doi.org/10.48550/arxiv.1805.10457, cornish2021heterodyned, Venumadhav2018} computing the real-time posterior distribution. However, the comparison with the optimized schemes is beyond the scope of the current work. In the follow-up works, we will compare our method against those schemes in detail.

Numerical experiments with a coherent, multi-detector implementation of the meshfree algorithm suggest that  we can solve the coherent BNS PE problem (over a ten-dimensional parameter space) in $\sim 164$ seconds using $64$ CPU cores.
Further optimizations and runs over the full parameter space are underway. These details will be available in a follow-up paper~\cite{pathak2023meshfree}.
It may be prudent to incorporate this method with the low-latency {\texttt{GstLAL}}~\cite{messick2017analysis} search pipeline for rapid, automated follow-ups of the detected events since both use the idea of dimension reduction using SVD.
In the current implementation of this algorithm, we need to create meshfree RBF interpolants from scratch for each new event triggered by the search pipelines. However, this task is embarrassingly parallel and can benefit from multiple CPU cores to expedite the preparatory stage. Future refinements could involve using the Fisher matrix~\cite{vallisneri2008use, pankow2015novel} as a guide to identifying the sample space volume and using sophisticated interpolation node distribution algorithms~\cite{Manca_2010}. Finally, the techniques of dimension reduction and meshfree approximation could be applied to situations where the likelihood function varies smoothly over the sample space. It is thus possible to adapt this idea to Bayesian inference in fields as diverse as cosmology, biochemical kinetic processes, and systems biology. 
\begin{acknowledgments}
We thank M.~K.~Haris for various suggestions and for his help at the conceptual stages of this work. We thank M.~Vallisneri and T.~Dent for carefully reading the manuscript and for offering several comments and suggestions to improve the presentation and content of the paper. We also thank other LSC colleagues S.~Mitra, A.~Ganguly, and S.~Kapadia, for their useful comments. L.~P. is supported by the Research Scholarship Program of Tata Consultancy Services (TCS). A.~R is supported by the research program of the Netherlands Organisation for Scientific Research (NWO). A.~S. gratefully acknowledges the generous grant provided by the Department of Science and Technology, India, through the DST-ICPS cluster project funding. We would like to thank the HPC support staff at IIT Gandhinagar for their help and cooperation. The authors are grateful for the computational resources provided by the LIGO Laboratory and supported by the National Science Foundation Grants No. PHY-0757058 and No. PHY-0823459. This material is based upon work supported by NSF's LIGO Laboratory, which is a major facility fully funded by the National Science Foundation. 
\end{acknowledgments}
\bibliography{references}

\begin{thebibliography}{36}%
\makeatletter
\providecommand \@ifxundefined [1]{%
 \@ifx{#1\undefined}
}%
\providecommand \@ifnum [1]{%
 \ifnum #1\expandafter \@firstoftwo
 \else \expandafter \@secondoftwo
 \fi
}%
\providecommand \@ifx [1]{%
 \ifx #1\expandafter \@firstoftwo
 \else \expandafter \@secondoftwo
 \fi
}%
\providecommand \natexlab [1]{#1}%
\providecommand \enquote  [1]{``#1''}%
\providecommand \bibnamefont  [1]{#1}%
\providecommand \bibfnamefont [1]{#1}%
\providecommand \citenamefont [1]{#1}%
\providecommand \href@noop [0]{\@secondoftwo}%
\providecommand \href [0]{\begingroup \@sanitize@url \@href}%
\providecommand \@href[1]{\@@startlink{#1}\@@href}%
\providecommand \@@href[1]{\endgroup#1\@@endlink}%
\providecommand \@sanitize@url [0]{\catcode `\\12\catcode `\$12\catcode `\&12\catcode `\#12\catcode `\^12\catcode `\_12\catcode `\%12\relax}%
\providecommand \@@startlink[1]{}%
\providecommand \@@endlink[0]{}%
\providecommand \url  [0]{\begingroup\@sanitize@url \@url }%
\providecommand \@url [1]{\endgroup\@href {#1}{\urlprefix }}%
\providecommand \urlprefix  [0]{URL }%
\providecommand \Eprint [0]{\href }%
\providecommand \doibase [0]{https://doi.org/}%
\providecommand \selectlanguage [0]{\@gobble}%
\providecommand \bibinfo  [0]{\@secondoftwo}%
\providecommand \bibfield  [0]{\@secondoftwo}%
\providecommand \translation [1]{[#1]}%
\providecommand \BibitemOpen [0]{}%
\providecommand \bibitemStop [0]{}%
\providecommand \bibitemNoStop [0]{.\EOS\space}%
\providecommand \EOS [0]{\spacefactor3000\relax}%
\providecommand \BibitemShut  [1]{\csname bibitem#1\endcsname}%
\let\auto@bib@innerbib\@empty
\bibitem [{\citenamefont {Abbott}\ \emph {et~al.}(2017)\citenamefont {Abbott} \emph {et~al.}}]{abbott2017gravitational}%
  \BibitemOpen
  \bibfield  {author} {\bibinfo {author} {\bibfnamefont {B.~P.}\ \bibnamefont {Abbott}} \emph {et~al.} (\bibinfo {collaboration} {{LIGO Scientific, Virgo, {\textit{Fermi}} Gamma-ray Burst Monitor and {\sc{INTEGRAL}} Collaboration}}),\ }\bibfield  {title} {\bibinfo {title} {Gravitational waves and gamma-rays from a binary neutron star merger: {GW170817} and {GRB 170817A}},\ }\href {https://doi.org/10.3847/2041-8213/aa920c} {\bibfield  {journal} {\bibinfo  {journal} {Astrophys. J. Lett.}\ }\textbf {\bibinfo {volume} {848}},\ \bibinfo {pages} {L13} (\bibinfo {year} {2017})}\BibitemShut {NoStop}%
\bibitem [{\citenamefont {Singer}\ and\ \citenamefont {Price}(2016)}]{singer2016rapid}%
  \BibitemOpen
  \bibfield  {author} {\bibinfo {author} {\bibfnamefont {L.~P.}\ \bibnamefont {Singer}}\ and\ \bibinfo {author} {\bibfnamefont {L.~R.}\ \bibnamefont {Price}},\ }\bibfield  {title} {\bibinfo {title} {Rapid {B}ayesian position reconstruction for gravitational-wave transients},\ }\href {https://doi.org/https://doi.org/10.1103/PhysRevD.93.024013} {\bibfield  {journal} {\bibinfo  {journal} {Phys. Rev. D}\ }\textbf {\bibinfo {volume} {93}},\ \bibinfo {pages} {024013} (\bibinfo {year} {2016})}\BibitemShut {NoStop}%
\bibitem [{\citenamefont {Finstad}\ and\ \citenamefont {Brown}(2020)}]{Finstad_2020}%
  \BibitemOpen
  \bibfield  {author} {\bibinfo {author} {\bibfnamefont {D.}~\bibnamefont {Finstad}}\ and\ \bibinfo {author} {\bibfnamefont {D.~A.}\ \bibnamefont {Brown}},\ }\bibfield  {title} {\bibinfo {title} {Fast parameter estimation of binary mergers for multimessenger follow-up},\ }\href {https://doi.org/10.3847/2041-8213/abca9e} {\bibfield  {journal} {\bibinfo  {journal} {Astrophys. J. Lett.}\ }\textbf {\bibinfo {volume} {905}},\ \bibinfo {pages} {L9} (\bibinfo {year} {2020})}\BibitemShut {NoStop}%
\bibitem [{\citenamefont {Abbott}\ \emph {et~al.}(2020)\citenamefont {Abbott} \emph {et~al.}}]{abbott2020prospects}%
  \BibitemOpen
  \bibfield  {author} {\bibinfo {author} {\bibfnamefont {B.~P.}\ \bibnamefont {Abbott}} \emph {et~al.} (\bibinfo {collaboration} {{KAGRA, LIGO Scientific and Virgo Collaboration}}),\ }\bibfield  {title} {\bibinfo {title} {{Prospects for observing and localizing gravitational-wave transients with Advanced LIGO, Advanced Virgo and KAGRA}},\ }\href {https://doi.org/10.1007/s41114-020-00026-9} {\bibfield  {journal} {\bibinfo  {journal} {Living Rev. Relativ.}\ }\textbf {\bibinfo {volume} {23}},\ \bibinfo {pages} {1} (\bibinfo {year} {2020})}\BibitemShut {NoStop}%
\bibitem [{\citenamefont {Margalit}\ and\ \citenamefont {Metzger}(2019)}]{Margalit_2019}%
  \BibitemOpen
  \bibfield  {author} {\bibinfo {author} {\bibfnamefont {B.}~\bibnamefont {Margalit}}\ and\ \bibinfo {author} {\bibfnamefont {B.~D.}\ \bibnamefont {Metzger}},\ }\bibfield  {title} {\bibinfo {title} {The multi-messenger matrix: The future of neutron star merger constraints on the nuclear equation of state},\ }\href {https://doi.org/10.3847/2041-8213/ab2ae2} {\bibfield  {journal} {\bibinfo  {journal} {Astrophys. J. Lett.}\ }\textbf {\bibinfo {volume} {880}},\ \bibinfo {pages} {L15} (\bibinfo {year} {2019})}\BibitemShut {NoStop}%
\bibitem [{\citenamefont {Dax}\ \emph {et~al.}(2021)\citenamefont {Dax}, \citenamefont {Green}, \citenamefont {Gair}, \citenamefont {Macke}, \citenamefont {Buonanno},\ and\ \citenamefont {Sch\"olkopf}}]{Dax2021}%
  \BibitemOpen
  \bibfield  {author} {\bibinfo {author} {\bibfnamefont {M.}~\bibnamefont {Dax}}, \bibinfo {author} {\bibfnamefont {S.~R.}\ \bibnamefont {Green}}, \bibinfo {author} {\bibfnamefont {J.}~\bibnamefont {Gair}}, \bibinfo {author} {\bibfnamefont {J.~H.}\ \bibnamefont {Macke}}, \bibinfo {author} {\bibfnamefont {A.}~\bibnamefont {Buonanno}},\ and\ \bibinfo {author} {\bibfnamefont {B.}~\bibnamefont {Sch\"olkopf}},\ }\bibfield  {title} {\bibinfo {title} {Real-time gravitational wave science with neural posterior estimation},\ }\href {https://doi.org/10.1103/PhysRevLett.127.241103} {\bibfield  {journal} {\bibinfo  {journal} {Phys. Rev. Lett.}\ }\textbf {\bibinfo {volume} {127}},\ \bibinfo {pages} {241103} (\bibinfo {year} {2021})}\BibitemShut {NoStop}%
\bibitem [{\citenamefont {Gabbard}\ \emph {et~al.}(2021)\citenamefont {Gabbard}, \citenamefont {Messenger}, \citenamefont {Heng}, \citenamefont {Tonolini},\ and\ \citenamefont {Murray-Smith}}]{gabbard2022bayesian}%
  \BibitemOpen
  \bibfield  {author} {\bibinfo {author} {\bibfnamefont {H.}~\bibnamefont {Gabbard}}, \bibinfo {author} {\bibfnamefont {C.}~\bibnamefont {Messenger}}, \bibinfo {author} {\bibfnamefont {I.~S.}\ \bibnamefont {Heng}}, \bibinfo {author} {\bibfnamefont {F.}~\bibnamefont {Tonolini}},\ and\ \bibinfo {author} {\bibfnamefont {R.}~\bibnamefont {Murray-Smith}},\ }\bibfield  {title} {\bibinfo {title} {Bayesian parameter estimation using conditional variational autoencoders for gravitational-wave astronomy},\ }\href {https://doi.org/10.1038/s41567-021-01425-7} {\bibfield  {journal} {\bibinfo  {journal} {Nature Physics}\ }\textbf {\bibinfo {volume} {18}},\ \bibinfo {pages} {112} (\bibinfo {year} {2021})}\BibitemShut {NoStop}%
\bibitem [{\citenamefont {Dax}\ \emph {et~al.}(2023)\citenamefont {Dax}, \citenamefont {Green}, \citenamefont {Gair}, \citenamefont {P\"urrer}, \citenamefont {Wildberger}, \citenamefont {Macke}, \citenamefont {Buonanno},\ and\ \citenamefont {Sch\"olkopf}}]{Dax2023}%
  \BibitemOpen
  \bibfield  {author} {\bibinfo {author} {\bibfnamefont {M.}~\bibnamefont {Dax}}, \bibinfo {author} {\bibfnamefont {S.~R.}\ \bibnamefont {Green}}, \bibinfo {author} {\bibfnamefont {J.}~\bibnamefont {Gair}}, \bibinfo {author} {\bibfnamefont {M.}~\bibnamefont {P\"urrer}}, \bibinfo {author} {\bibfnamefont {J.}~\bibnamefont {Wildberger}}, \bibinfo {author} {\bibfnamefont {J.~H.}\ \bibnamefont {Macke}}, \bibinfo {author} {\bibfnamefont {A.}~\bibnamefont {Buonanno}},\ and\ \bibinfo {author} {\bibfnamefont {B.}~\bibnamefont {Sch\"olkopf}},\ }\bibfield  {title} {\bibinfo {title} {Neural importance sampling for rapid and reliable gravitational-wave inference},\ }\href {https://doi.org/10.1103/PhysRevLett.130.171403} {\bibfield  {journal} {\bibinfo  {journal} {Phys. Rev. Lett.}\ }\textbf {\bibinfo {volume} {130}},\ \bibinfo {pages} {171403} (\bibinfo {year} {2023})}\BibitemShut {NoStop}%
\bibitem [{\citenamefont {Canizares}\ \emph {et~al.}(2015)\citenamefont {Canizares}, \citenamefont {Field}, \citenamefont {Gair}, \citenamefont {Raymond}, \citenamefont {Smith},\ and\ \citenamefont {Tiglio}}]{canizares2015accelerated}%
  \BibitemOpen
  \bibfield  {author} {\bibinfo {author} {\bibfnamefont {P.}~\bibnamefont {Canizares}}, \bibinfo {author} {\bibfnamefont {S.~E.}\ \bibnamefont {Field}}, \bibinfo {author} {\bibfnamefont {J.}~\bibnamefont {Gair}}, \bibinfo {author} {\bibfnamefont {V.}~\bibnamefont {Raymond}}, \bibinfo {author} {\bibfnamefont {R.}~\bibnamefont {Smith}},\ and\ \bibinfo {author} {\bibfnamefont {M.}~\bibnamefont {Tiglio}},\ }\bibfield  {title} {\bibinfo {title} {Accelerated gravitational wave parameter estimation with reduced order modeling},\ }\href {https://doi.org/https://doi.org/10.1103/PhysRevLett.114.071104} {\bibfield  {journal} {\bibinfo  {journal} {Phy. Rev. Lett.}\ }\textbf {\bibinfo {volume} {114}},\ \bibinfo {pages} {071104} (\bibinfo {year} {2015})}\BibitemShut {NoStop}%
\bibitem [{\citenamefont {Qi}\ and\ \citenamefont {Raymond}(2021)}]{Qi_2021}%
  \BibitemOpen
  \bibfield  {author} {\bibinfo {author} {\bibfnamefont {H.}~\bibnamefont {Qi}}\ and\ \bibinfo {author} {\bibfnamefont {V.}~\bibnamefont {Raymond}},\ }\bibfield  {title} {\bibinfo {title} {Python-based reduced order quadrature building code for fast gravitational wave inference},\ }\bibfield  {journal} {\bibinfo  {journal} {Phys. Rev. D}\ }\textbf {\bibinfo {volume} {104}},\ \href {https://doi.org/10.1103/physrevd.104.063031} {10.1103/physrevd.104.063031} (\bibinfo {year} {2021})\BibitemShut {NoStop}%
\bibitem [{\citenamefont {Morisaki}\ and\ \citenamefont {Raymond}(2020)}]{Soichiro_2020}%
  \BibitemOpen
  \bibfield  {author} {\bibinfo {author} {\bibfnamefont {S.}~\bibnamefont {Morisaki}}\ and\ \bibinfo {author} {\bibfnamefont {V.}~\bibnamefont {Raymond}},\ }\bibfield  {title} {\bibinfo {title} {Rapid parameter estimation of gravitational waves from binary neutron star coalescence using focused reduced order quadrature},\ }\href {https://doi.org/10.1103/PhysRevD.102.104020} {\bibfield  {journal} {\bibinfo  {journal} {Phys. Rev. D}\ }\textbf {\bibinfo {volume} {102}},\ \bibinfo {pages} {104020} (\bibinfo {year} {2020})}\BibitemShut {NoStop}%
\bibitem [{\citenamefont {Chua}\ and\ \citenamefont {Vallisneri}(2020)}]{chua2020learning}%
  \BibitemOpen
  \bibfield  {author} {\bibinfo {author} {\bibfnamefont {A.~J.}\ \bibnamefont {Chua}}\ and\ \bibinfo {author} {\bibfnamefont {M.}~\bibnamefont {Vallisneri}},\ }\bibfield  {title} {\bibinfo {title} {Learning {B}ayesian posteriors with neural networks for gravitational-wave inference},\ }\href {https://doi.org/https://doi.org/10.1103/PhysRevLett.124.041102} {\bibfield  {journal} {\bibinfo  {journal} {Phys. Rev. Lett.}\ }\textbf {\bibinfo {volume} {124}},\ \bibinfo {pages} {041102} (\bibinfo {year} {2020})}\BibitemShut {NoStop}%
\bibitem [{\citenamefont {Lange}\ \emph {et~al.}(2018)\citenamefont {Lange}, \citenamefont {O'Shaughnessy},\ and\ \citenamefont {Rizzo}}]{https://doi.org/10.48550/arxiv.1805.10457}%
  \BibitemOpen
  \bibfield  {author} {\bibinfo {author} {\bibfnamefont {J.}~\bibnamefont {Lange}}, \bibinfo {author} {\bibfnamefont {R.}~\bibnamefont {O'Shaughnessy}},\ and\ \bibinfo {author} {\bibfnamefont {M.}~\bibnamefont {Rizzo}},\ }\href {https://doi.org/10.48550/ARXIV.1805.10457} {\bibinfo {title} {Rapid and accurate parameter inference for coalescing, precessing compact binaries}} (\bibinfo {year} {2018})\BibitemShut {NoStop}%
\bibitem [{\citenamefont {Cornish}(2021)}]{cornish2021heterodyned}%
  \BibitemOpen
  \bibfield  {author} {\bibinfo {author} {\bibfnamefont {N.~J.}\ \bibnamefont {Cornish}},\ }\bibfield  {title} {\bibinfo {title} {Heterodyned likelihood for rapid gravitational wave parameter inference},\ }\href {https://doi.org/10.1103/PhysRevD.104.104054} {\bibfield  {journal} {\bibinfo  {journal} {Phys. Rev. D}\ }\textbf {\bibinfo {volume} {104}},\ \bibinfo {pages} {104054} (\bibinfo {year} {2021})}\BibitemShut {NoStop}%
\bibitem [{\citenamefont {Zackay}\ \emph {et~al.}(2018)\citenamefont {Zackay}, \citenamefont {Dai},\ and\ \citenamefont {Venumadhav}}]{Venumadhav2018}%
  \BibitemOpen
  \bibfield  {author} {\bibinfo {author} {\bibfnamefont {B.}~\bibnamefont {Zackay}}, \bibinfo {author} {\bibfnamefont {L.}~\bibnamefont {Dai}},\ and\ \bibinfo {author} {\bibfnamefont {T.}~\bibnamefont {Venumadhav}},\ }\href {https://doi.org/10.48550/ARXIV.1806.08792} {\bibinfo {title} {Relative binning and fast likelihood evaluation for gravitational wave parameter estimation}} (\bibinfo {year} {2018}),\ \Eprint {https://arxiv.org/abs/1806.08792} {arXiv:1806.08792 [astro-ph.IM]} \BibitemShut {NoStop}%
\bibitem [{\citenamefont {Smith}\ \emph {et~al.}(2014)\citenamefont {Smith}, \citenamefont {Hanna}, \citenamefont {Mandel},\ and\ \citenamefont {Vecchio}}]{smith2014rapidly}%
  \BibitemOpen
  \bibfield  {author} {\bibinfo {author} {\bibfnamefont {R.}~\bibnamefont {Smith}}, \bibinfo {author} {\bibfnamefont {C.}~\bibnamefont {Hanna}}, \bibinfo {author} {\bibfnamefont {I.}~\bibnamefont {Mandel}},\ and\ \bibinfo {author} {\bibfnamefont {A.}~\bibnamefont {Vecchio}},\ }\bibfield  {title} {\bibinfo {title} {Rapidly evaluating the compact-binary likelihood function via interpolation},\ }\href {https://doi.org/https://doi.org/10.1103/PhysRevD.90.044074} {\bibfield  {journal} {\bibinfo  {journal} {Phys. Rev. D}\ }\textbf {\bibinfo {volume} {90}},\ \bibinfo {pages} {044074} (\bibinfo {year} {2014})}\BibitemShut {NoStop}%
\bibitem [{\citenamefont {Cannon}\ \emph {et~al.}(2010)\citenamefont {Cannon}, \citenamefont {Chapman}, \citenamefont {Hanna}, \citenamefont {Keppel}, \citenamefont {Searle},\ and\ \citenamefont {Weinstein}}]{GstLAL_2010}%
  \BibitemOpen
  \bibfield  {author} {\bibinfo {author} {\bibfnamefont {K.}~\bibnamefont {Cannon}}, \bibinfo {author} {\bibfnamefont {A.}~\bibnamefont {Chapman}}, \bibinfo {author} {\bibfnamefont {C.}~\bibnamefont {Hanna}}, \bibinfo {author} {\bibfnamefont {D.}~\bibnamefont {Keppel}}, \bibinfo {author} {\bibfnamefont {A.~C.}\ \bibnamefont {Searle}},\ and\ \bibinfo {author} {\bibfnamefont {A.~J.}\ \bibnamefont {Weinstein}},\ }\bibfield  {title} {\bibinfo {title} {Singular value decomposition applied to compact binary coalescence gravitational-wave signals},\ }\href {https://doi.org/10.1103/PhysRevD.82.044025} {\bibfield  {journal} {\bibinfo  {journal} {Phys. Rev. D}\ }\textbf {\bibinfo {volume} {82}},\ \bibinfo {pages} {044025} (\bibinfo {year} {2010})}\BibitemShut {NoStop}%
\bibitem [{\citenamefont {Golub}\ and\ \citenamefont {Van~Loan}(2013)}]{golub2013matrix}%
  \BibitemOpen
  \bibfield  {author} {\bibinfo {author} {\bibfnamefont {G.~H.}\ \bibnamefont {Golub}}\ and\ \bibinfo {author} {\bibfnamefont {C.~F.}\ \bibnamefont {Van~Loan}},\ }\href {https://www.press.jhu.edu/books/title/10678/matrix-computations} {\emph {\bibinfo {title} {Matrix Computations}}},\ \bibinfo {edition} {{Fourth}}\ ed.\ (\bibinfo  {publisher} {The Johns Hopkins University Press},\ \bibinfo {year} {2013})\BibitemShut {NoStop}%
\bibitem [{\citenamefont {Foreman-Mackey}\ \emph {et~al.}(2013)\citenamefont {Foreman-Mackey}, \citenamefont {Hogg}, \citenamefont {Lang},\ and\ \citenamefont {Goodman}}]{Foreman_Mackey_2013}%
  \BibitemOpen
  \bibfield  {author} {\bibinfo {author} {\bibfnamefont {D.}~\bibnamefont {Foreman-Mackey}}, \bibinfo {author} {\bibfnamefont {D.~W.}\ \bibnamefont {Hogg}}, \bibinfo {author} {\bibfnamefont {D.}~\bibnamefont {Lang}},\ and\ \bibinfo {author} {\bibfnamefont {J.}~\bibnamefont {Goodman}},\ }\bibfield  {title} {\bibinfo {title} {{\texttt{emcee}}: The {MCMC} {H}ammer},\ }\href {https://doi.org/10.1086/670067} {\bibfield  {journal} {\bibinfo  {journal} {Publ. Astron. Soc. Pac.}\ }\textbf {\bibinfo {volume} {125}},\ \bibinfo {pages} {306} (\bibinfo {year} {2013})}\BibitemShut {NoStop}%
\bibitem [{\citenamefont {Allen}\ \emph {et~al.}(2012)\citenamefont {Allen}, \citenamefont {Anderson}, \citenamefont {Brady}, \citenamefont {Brown},\ and\ \citenamefont {Creighton}}]{findchirp_2012}%
  \BibitemOpen
  \bibfield  {author} {\bibinfo {author} {\bibfnamefont {B.}~\bibnamefont {Allen}}, \bibinfo {author} {\bibfnamefont {W.~G.}\ \bibnamefont {Anderson}}, \bibinfo {author} {\bibfnamefont {P.~R.}\ \bibnamefont {Brady}}, \bibinfo {author} {\bibfnamefont {D.~A.}\ \bibnamefont {Brown}},\ and\ \bibinfo {author} {\bibfnamefont {J.~D.~E.}\ \bibnamefont {Creighton}},\ }\bibfield  {title} {\bibinfo {title} {Findchirp: An algorithm for detection of gravitational waves from inspiraling compact binaries},\ }\href {https://doi.org/10.1103/PhysRevD.85.122006} {\bibfield  {journal} {\bibinfo  {journal} {Phys. Rev. D}\ }\textbf {\bibinfo {volume} {85}},\ \bibinfo {pages} {122006} (\bibinfo {year} {2012})}\BibitemShut {NoStop}%
\bibitem [{\citenamefont {Skilling}(2006)}]{skilling2006nested}%
  \BibitemOpen
  \bibfield  {author} {\bibinfo {author} {\bibfnamefont {J.}~\bibnamefont {Skilling}},\ }\bibfield  {title} {\bibinfo {title} {{Nested sampling for general Bayesian computation}},\ }\href {https://doi.org/10.1214/06-BA127} {\bibfield  {journal} {\bibinfo  {journal} {{Bayesian Anal.}}\ }\textbf {\bibinfo {volume} {1}},\ \bibinfo {pages} {833} (\bibinfo {year} {2006})}\BibitemShut {NoStop}%
\bibitem [{\citenamefont {Thrane}\ and\ \citenamefont {Talbot}(2019)}]{thrane_2019}%
  \BibitemOpen
  \bibfield  {author} {\bibinfo {author} {\bibfnamefont {E.}~\bibnamefont {Thrane}}\ and\ \bibinfo {author} {\bibfnamefont {C.}~\bibnamefont {Talbot}},\ }\bibfield  {title} {\bibinfo {title} {An introduction to {Bayesian} inference in gravitational-wave astronomy: parameter estimation, model selection, and hierarchical models},\ }\bibfield  {journal} {\bibinfo  {journal} {Publications of the Astronomical Society of Australia}\ }\textbf {\bibinfo {volume} {36}},\ \href {https://doi.org/https://doi.org/10.1017/pasa.2019.2} {https://doi.org/10.1017/pasa.2019.2} (\bibinfo {year} {2019})\BibitemShut {NoStop}%
\bibitem [{\citenamefont {Messick}\ \emph {et~al.}(2017)\citenamefont {Messick}, \citenamefont {Blackburn}, \citenamefont {Brady}, \citenamefont {Brockill}, \citenamefont {Cannon}, \citenamefont {Cariou}, \citenamefont {Caudill}, \citenamefont {Chamberlin}, \citenamefont {Creighton}, \citenamefont {Everett} \emph {et~al.}}]{messick2017analysis}%
  \BibitemOpen
  \bibfield  {author} {\bibinfo {author} {\bibfnamefont {C.}~\bibnamefont {Messick}}, \bibinfo {author} {\bibfnamefont {K.}~\bibnamefont {Blackburn}}, \bibinfo {author} {\bibfnamefont {P.}~\bibnamefont {Brady}}, \bibinfo {author} {\bibfnamefont {P.}~\bibnamefont {Brockill}}, \bibinfo {author} {\bibfnamefont {K.}~\bibnamefont {Cannon}}, \bibinfo {author} {\bibfnamefont {R.}~\bibnamefont {Cariou}}, \bibinfo {author} {\bibfnamefont {S.}~\bibnamefont {Caudill}}, \bibinfo {author} {\bibfnamefont {S.~J.}\ \bibnamefont {Chamberlin}}, \bibinfo {author} {\bibfnamefont {J.~D.}\ \bibnamefont {Creighton}}, \bibinfo {author} {\bibfnamefont {R.}~\bibnamefont {Everett}}, \emph {et~al.},\ }\bibfield  {title} {\bibinfo {title} {Analysis framework for the prompt discovery of compact binary mergers in gravitational-wave data},\ }\href {https://doi.org/https://doi.org/10.1103/PhysRevD.95.042001} {\bibfield  {journal} {\bibinfo  {journal} {Phys. Rev. D}\ }\textbf {\bibinfo {volume} {95}},\ \bibinfo {pages} {042001} (\bibinfo
  {year} {2017})}\BibitemShut {NoStop}%
\bibitem [{\citenamefont {Usman}\ \emph {et~al.}(2016)\citenamefont {Usman}, \citenamefont {Nitz}, \citenamefont {Harry}, \citenamefont {Biwer}, \citenamefont {Brown}, \citenamefont {Cabero}, \citenamefont {Capano}, \citenamefont {Dal~Canton}, \citenamefont {Dent}, \citenamefont {Fairhurst} \emph {et~al.}}]{usman2016pycbc}%
  \BibitemOpen
  \bibfield  {author} {\bibinfo {author} {\bibfnamefont {S.~A.}\ \bibnamefont {Usman}}, \bibinfo {author} {\bibfnamefont {A.~H.}\ \bibnamefont {Nitz}}, \bibinfo {author} {\bibfnamefont {I.~W.}\ \bibnamefont {Harry}}, \bibinfo {author} {\bibfnamefont {C.~M.}\ \bibnamefont {Biwer}}, \bibinfo {author} {\bibfnamefont {D.~A.}\ \bibnamefont {Brown}}, \bibinfo {author} {\bibfnamefont {M.}~\bibnamefont {Cabero}}, \bibinfo {author} {\bibfnamefont {C.~D.}\ \bibnamefont {Capano}}, \bibinfo {author} {\bibfnamefont {T.}~\bibnamefont {Dal~Canton}}, \bibinfo {author} {\bibfnamefont {T.}~\bibnamefont {Dent}}, \bibinfo {author} {\bibfnamefont {S.}~\bibnamefont {Fairhurst}}, \emph {et~al.},\ }\bibfield  {title} {\bibinfo {title} {The {PyCBC} search for gravitational waves from compact binary coalescence},\ }\href {https://doi.org/10.1088/0264-9381/33/21/215004} {\bibfield  {journal} {\bibinfo  {journal} {Class. Quantum Gravity}\ }\textbf {\bibinfo {volume} {33}},\ \bibinfo {pages} {215004} (\bibinfo {year} {2016})}\BibitemShut
  {NoStop}%
\bibitem [{\citenamefont {Fasshauer}(2007)}]{doi:10.1142/6437}%
  \BibitemOpen
  \bibfield  {author} {\bibinfo {author} {\bibfnamefont {G.~E.}\ \bibnamefont {Fasshauer}},\ }\href {https://doi.org/https://doi.org/10.1142/6437} {\emph {\bibinfo {title} {{Meshfree Approximation Methods with Matlab}}}}\ (\bibinfo  {publisher} {World Scientific},\ \bibinfo {year} {2007})\BibitemShut {NoStop}%
\bibitem [{\citenamefont {Khan}\ \emph {et~al.}(2016)\citenamefont {Khan}, \citenamefont {Husa}, \citenamefont {Hannam}, \citenamefont {Ohme}, \citenamefont {P{\"u}rrer}, \citenamefont {Forteza},\ and\ \citenamefont {Boh{\'e}}}]{khan2016frequency}%
  \BibitemOpen
  \bibfield  {author} {\bibinfo {author} {\bibfnamefont {S.}~\bibnamefont {Khan}}, \bibinfo {author} {\bibfnamefont {S.}~\bibnamefont {Husa}}, \bibinfo {author} {\bibfnamefont {M.}~\bibnamefont {Hannam}}, \bibinfo {author} {\bibfnamefont {F.}~\bibnamefont {Ohme}}, \bibinfo {author} {\bibfnamefont {M.}~\bibnamefont {P{\"u}rrer}}, \bibinfo {author} {\bibfnamefont {X.~J.}\ \bibnamefont {Forteza}},\ and\ \bibinfo {author} {\bibfnamefont {A.}~\bibnamefont {Boh{\'e}}},\ }\bibfield  {title} {\bibinfo {title} {{Frequency-domain gravitational waves from nonprecessing black-hole binaries. {II.} A phenomenological model for the advanced detector era}},\ }\href {https://doi.org/10.1103/PhysRevD.93.044007} {\bibfield  {journal} {\bibinfo  {journal} {Phys. Rev. D}\ }\textbf {\bibinfo {volume} {93}},\ \bibinfo {pages} {044007} (\bibinfo {year} {2016})}\BibitemShut {NoStop}%
\bibitem [{\citenamefont {Barsotti}\ \emph {et~al.}(2018)\citenamefont {Barsotti}, \citenamefont {Gras}, \citenamefont {Evans},\ and\ \citenamefont {Fritschel}}]{aLIGO_ZDHP}%
  \BibitemOpen
  \bibfield  {author} {\bibinfo {author} {\bibfnamefont {L.}~\bibnamefont {Barsotti}}, \bibinfo {author} {\bibfnamefont {S.}~\bibnamefont {Gras}}, \bibinfo {author} {\bibfnamefont {M.}~\bibnamefont {Evans}},\ and\ \bibinfo {author} {\bibfnamefont {P.}~\bibnamefont {Fritschel}},\ }\href {https://dcc.ligo.org/LIGO-T1800044/public} {\emph {\bibinfo {title} {{The updated Advanced LIGO design curve}}}},\ \bibinfo {type} {Tech. Rep.}\ \bibinfo {number} {LIGO-T1800044-v5}\ (\bibinfo  {institution} {LIGO Scientific Collaboration},\ \bibinfo {year} {2018})\BibitemShut {NoStop}%
\bibitem [{\citenamefont {Hines}(2015)}]{RBF_github}%
  \BibitemOpen
  \bibfield  {author} {\bibinfo {author} {\bibfnamefont {T.}~\bibnamefont {Hines}},\ }\href {https://github.com/treverhines/RBF.git} {\bibinfo {title} {Python package containing the tools necessary for radial basis function {(RBF)} applications}} (\bibinfo {year} {2015})\BibitemShut {NoStop}%
\bibitem [{\citenamefont {Speagle}(2020)}]{speagle2020dynesty}%
  \BibitemOpen
  \bibfield  {author} {\bibinfo {author} {\bibfnamefont {J.~S.}\ \bibnamefont {Speagle}},\ }\bibfield  {title} {\bibinfo {title} {{\sc{DYNESTY}}: a dynamic nested sampling package for estimating {Bayesian} posteriors and evidences},\ }\href {https://doi.org/10.1093/mnras/staa278} {\bibfield  {journal} {\bibinfo  {journal} {Mon. Not. R. Astron. Soc.}\ }\textbf {\bibinfo {volume} {493}},\ \bibinfo {pages} {3132} (\bibinfo {year} {2020})}\BibitemShut {NoStop}%
\bibitem [{\citenamefont {Koposov}\ \emph {et~al.}(2023)\citenamefont {Koposov}, \citenamefont {Speagle}, \citenamefont {Barbary}, \citenamefont {Ashton}, \citenamefont {Bennett}, \citenamefont {Buchner}, \citenamefont {Scheffler}, \citenamefont {Cook}, \citenamefont {Talbot}, \citenamefont {Guillochon} \emph {et~al.}}]{sergey_koposov_2023_7600689}%
  \BibitemOpen
  \bibfield  {author} {\bibinfo {author} {\bibfnamefont {S.}~\bibnamefont {Koposov}}, \bibinfo {author} {\bibfnamefont {J.}~\bibnamefont {Speagle}}, \bibinfo {author} {\bibfnamefont {K.}~\bibnamefont {Barbary}}, \bibinfo {author} {\bibfnamefont {G.}~\bibnamefont {Ashton}}, \bibinfo {author} {\bibfnamefont {E.}~\bibnamefont {Bennett}}, \bibinfo {author} {\bibfnamefont {J.}~\bibnamefont {Buchner}}, \bibinfo {author} {\bibfnamefont {C.}~\bibnamefont {Scheffler}}, \bibinfo {author} {\bibfnamefont {B.}~\bibnamefont {Cook}}, \bibinfo {author} {\bibfnamefont {C.}~\bibnamefont {Talbot}}, \bibinfo {author} {\bibfnamefont {J.}~\bibnamefont {Guillochon}}, \emph {et~al.},\ }\href {https://doi.org/10.5281/zenodo.7600689} {\bibinfo {title} {joshspeagle/dynesty: v2.1.0}} (\bibinfo {year} {2023})\BibitemShut {NoStop}%
\bibitem [{\citenamefont {Ayanendranath~Basu}(2011)}]{stats_book}%
  \BibitemOpen
  \bibfield  {author} {\bibinfo {author} {\bibfnamefont {C.~P.}\ \bibnamefont {Ayanendranath~Basu}, \bibfnamefont {Hiroyuki~Shioya}},\ }\href {https://doi.org/https://doi.org/10.1201/b10956} {\emph {\bibinfo {title} {{Statistical Inference: The Minimum Distance Approach}}}},\ \bibinfo {edition} {1st}\ ed.\ (\bibinfo  {publisher} {Chapman and Hall/CRC},\ \bibinfo {year} {2011})\BibitemShut {NoStop}%
\bibitem [{\citenamefont {Biwer}\ \emph {et~al.}(2019)\citenamefont {Biwer}, \citenamefont {Capano}, \citenamefont {De}, \citenamefont {Cabero}, \citenamefont {Brown}, \citenamefont {Nitz},\ and\ \citenamefont {Raymond}}]{biwer2019pycbc}%
  \BibitemOpen
  \bibfield  {author} {\bibinfo {author} {\bibfnamefont {C.~M.}\ \bibnamefont {Biwer}}, \bibinfo {author} {\bibfnamefont {C.~D.}\ \bibnamefont {Capano}}, \bibinfo {author} {\bibfnamefont {S.}~\bibnamefont {De}}, \bibinfo {author} {\bibfnamefont {M.}~\bibnamefont {Cabero}}, \bibinfo {author} {\bibfnamefont {D.~A.}\ \bibnamefont {Brown}}, \bibinfo {author} {\bibfnamefont {A.~H.}\ \bibnamefont {Nitz}},\ and\ \bibinfo {author} {\bibfnamefont {V.}~\bibnamefont {Raymond}},\ }\bibfield  {title} {\bibinfo {title} {{PyCBC Inference:} a {Python-based} parameter estimation toolkit for compact binary coalescence signals},\ }\href {https://doi.org/10.1088/1538-3873/aaef0b} {\bibfield  {journal} {\bibinfo  {journal} {Publ. Astron. Soc. Pac.}\ }\textbf {\bibinfo {volume} {131}},\ \bibinfo {pages} {024503} (\bibinfo {year} {2019})}\BibitemShut {NoStop}%
\bibitem [{\citenamefont {Pathak}\ \emph {et~al.}(2023)\citenamefont {Pathak}, \citenamefont {Munishwar}, \citenamefont {Reza},\ and\ \citenamefont {Sengupta}}]{pathak2023meshfree}%
  \BibitemOpen
  \bibfield  {author} {\bibinfo {author} {\bibfnamefont {L.}~\bibnamefont {Pathak}}, \bibinfo {author} {\bibfnamefont {S.}~\bibnamefont {Munishwar}}, \bibinfo {author} {\bibfnamefont {A.}~\bibnamefont {Reza}},\ and\ \bibinfo {author} {\bibfnamefont {A.~S.}\ \bibnamefont {Sengupta}},\ }\href@noop {} {\bibinfo {title} {Prompt sky localization of {EM} counterparts of {GW} transients using meshfree approximations}} (\bibinfo {year} {in preparation (2023)})\BibitemShut {NoStop}%
\bibitem [{\citenamefont {Vallisneri}(2008)}]{vallisneri2008use}%
  \BibitemOpen
  \bibfield  {author} {\bibinfo {author} {\bibfnamefont {M.}~\bibnamefont {Vallisneri}},\ }\bibfield  {title} {\bibinfo {title} {Use and abuse of the {Fisher} information matrix in the assessment of gravitational-wave parameter-estimation prospects},\ }\href {https://doi.org/10.1103/PhysRevD.77.042001} {\bibfield  {journal} {\bibinfo  {journal} {Phys. Rev. D}\ }\textbf {\bibinfo {volume} {77}},\ \bibinfo {pages} {042001} (\bibinfo {year} {2008})}\BibitemShut {NoStop}%
\bibitem [{\citenamefont {Pankow}\ \emph {et~al.}(2015)\citenamefont {Pankow}, \citenamefont {Brady}, \citenamefont {Ochsner},\ and\ \citenamefont {O'Shaughnessy}}]{pankow2015novel}%
  \BibitemOpen
  \bibfield  {author} {\bibinfo {author} {\bibfnamefont {C.}~\bibnamefont {Pankow}}, \bibinfo {author} {\bibfnamefont {P.}~\bibnamefont {Brady}}, \bibinfo {author} {\bibfnamefont {E.}~\bibnamefont {Ochsner}},\ and\ \bibinfo {author} {\bibfnamefont {R.}~\bibnamefont {O'Shaughnessy}},\ }\bibfield  {title} {\bibinfo {title} {Novel scheme for rapid parallel parameter estimation of gravitational waves from compact binary coalescences},\ }\href {https://doi.org/10.1103/PhysRevD.92.023002} {\bibfield  {journal} {\bibinfo  {journal} {Phys. Rev. D}\ }\textbf {\bibinfo {volume} {92}},\ \bibinfo {pages} {023002} (\bibinfo {year} {2015})}\BibitemShut {NoStop}%
\bibitem [{\citenamefont {Manca}\ and\ \citenamefont {Vallisneri}(2010)}]{Manca_2010}%
  \BibitemOpen
  \bibfield  {author} {\bibinfo {author} {\bibfnamefont {G.~M.}\ \bibnamefont {Manca}}\ and\ \bibinfo {author} {\bibfnamefont {M.}~\bibnamefont {Vallisneri}},\ }\bibfield  {title} {\bibinfo {title} {Cover art: Issues in the metric-guided and metric-less placement of random and stochastic template banks},\ }\href {https://doi.org/10.1103/PhysRevD.81.024004} {\bibfield  {journal} {\bibinfo  {journal} {Phys. Rev. D}\ }\textbf {\bibinfo {volume} {81}},\ \bibinfo {pages} {024004} (\bibinfo {year} {2010})}\BibitemShut {NoStop}%
\end{thebibliography}%

\appendix
\section{ACCURACY TRADE-OFFS: EFFECT OF VARYING BASIS SIZE}
\label{appendix:A}
\begin{figure}[htb!]
\includegraphics[width = \columnwidth, clip = True]{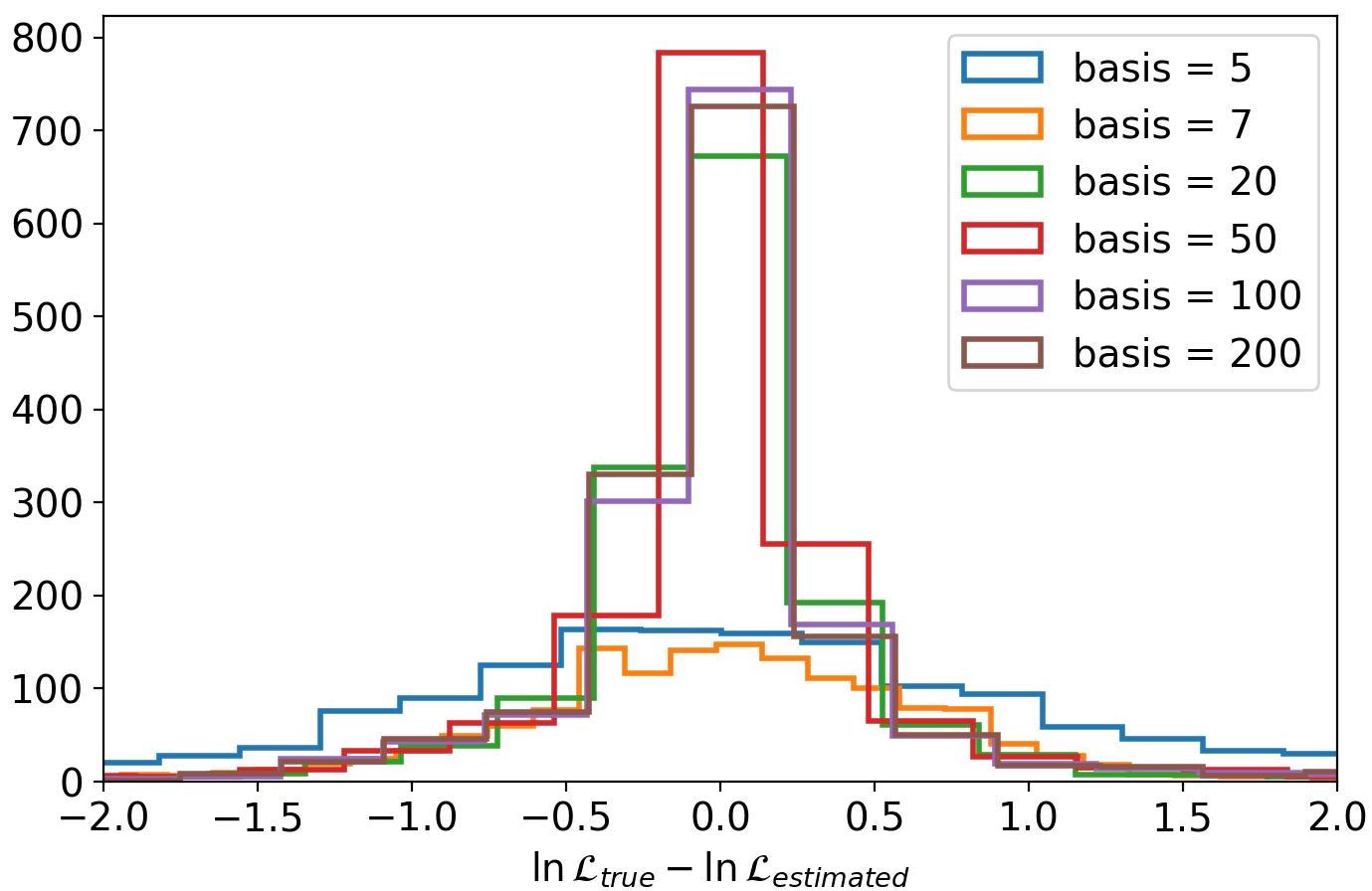}
\caption{The likelihood reconstruction error has been shown by varying the number of basis vectors. 
}
\label{Fig:Error_with_basis_pdf}
\end{figure}

\begin{figure}[htb!]
\includegraphics[width = \columnwidth, clip = True]{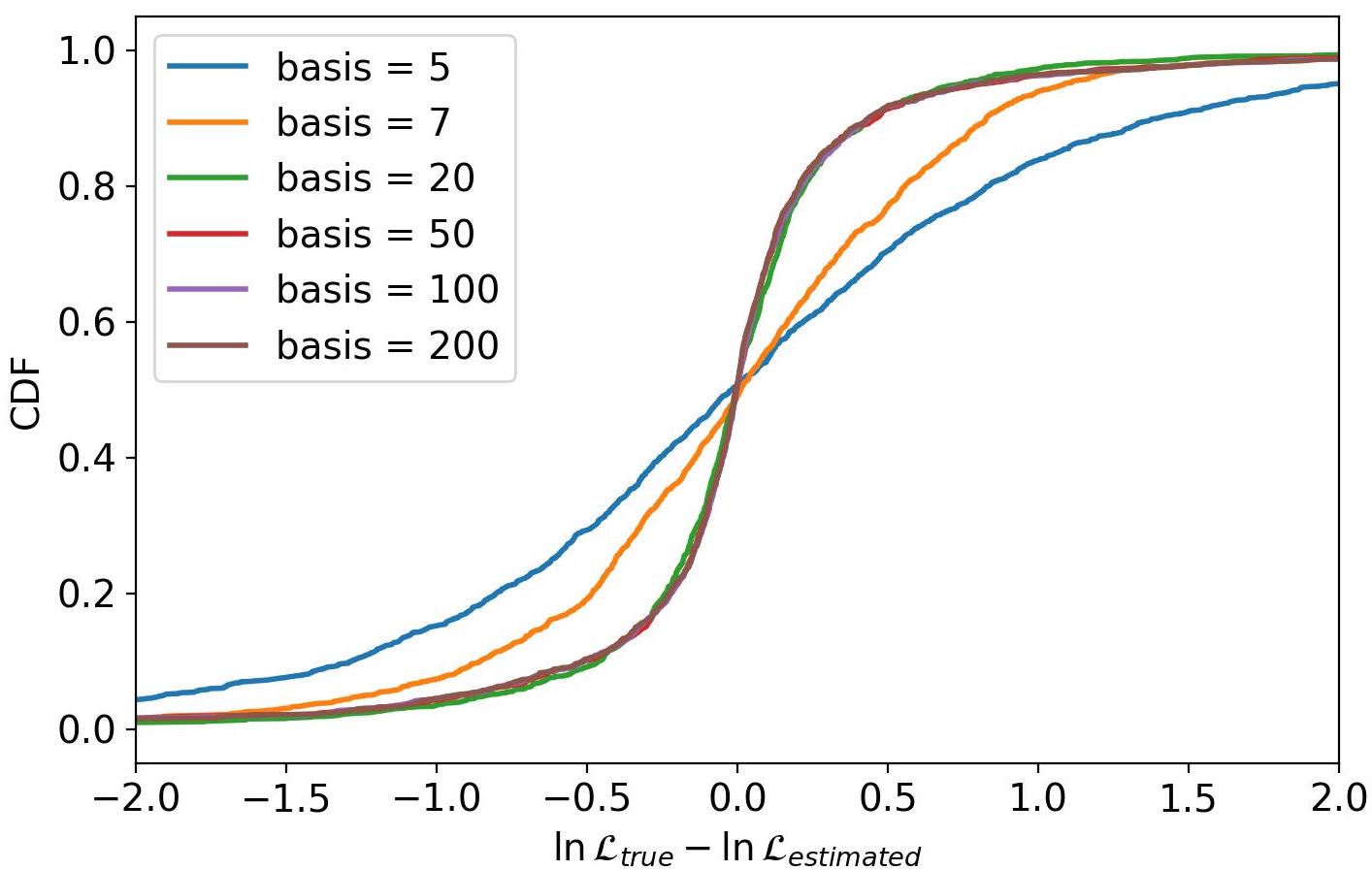}
\caption{The CDFs with different number of basis vectors has been plotted for the same likelihood reconstruction error shown in Fig. \ref{Fig:Error_with_basis_pdf}. }
\label{Fig:Error_with_basis_cdf}
\end{figure}

We present the results of a simulation to demonstrate the accuracy of the mesh-free method on the basis size. We inject  a GW signal from a compact binary system with component masses ($1.4 M_{\odot}$, $1.4 M_{\odot}$) in simulated noise generated using the  advanced LIGO design sensitivity curve with a lower seismic cut-off frequency of $10$ Hz. 
On one hand, we compute the log-likelihood-ratio (LLR) using the standard functions available in the PyCBC package (considered to be the ‘ground truth’ for this simulation) and compare it with the approximate value using the mesh-free interpolation method as outlined in our manuscript. 
The meshfree method uses $10^{3}$ randomly chosen initial nodes over the sample space. We vary the number of the basis vectors for reconstructing the likelihood and estimate the difference between interpolated likelihood and the ‘ground-truth’ likelihood value. 
For these simulations, we observe that the top $20$ basis vectors are sufficient to approximate the likelihood with sufficient accuracy. However, we demonstrated the absolute error between the true and approximated value for $5$, $7$, $20$, $50$, $100$, and $200$ basis vectors, respectively. Fig.~\ref{Fig:Error_with_basis_pdf} shows the probability distribution function (PDF) of the difference between true and estimated likelihood, and the Fig.~\ref{Fig:Error_with_basis_cdf} shows the corresponding cumulative distribution function (CDF). It is clear from these figures that as we increase the number of basis vectors, the error distribution becomes more concentrated around zero as compared to the relatively flattened distribution for smaller basis sizes.
\section{SCALING OF THE SPEED-UP FACTOR WITH WAVEFORM DURATION}
The meshfree likelihood evaluation has no dependence on the length of the waveform. On the other hand, in the standard method, a significant time is spent on the waveform generation, followed by the evaluation of the likelihood integral. The latter depends on the length of the waveform. As shown in Table II, the speed-up ratios decrease with higher masses (shorter waveforms). 

\begin{figure}[htb!]
\includegraphics[width = \columnwidth, clip = True]{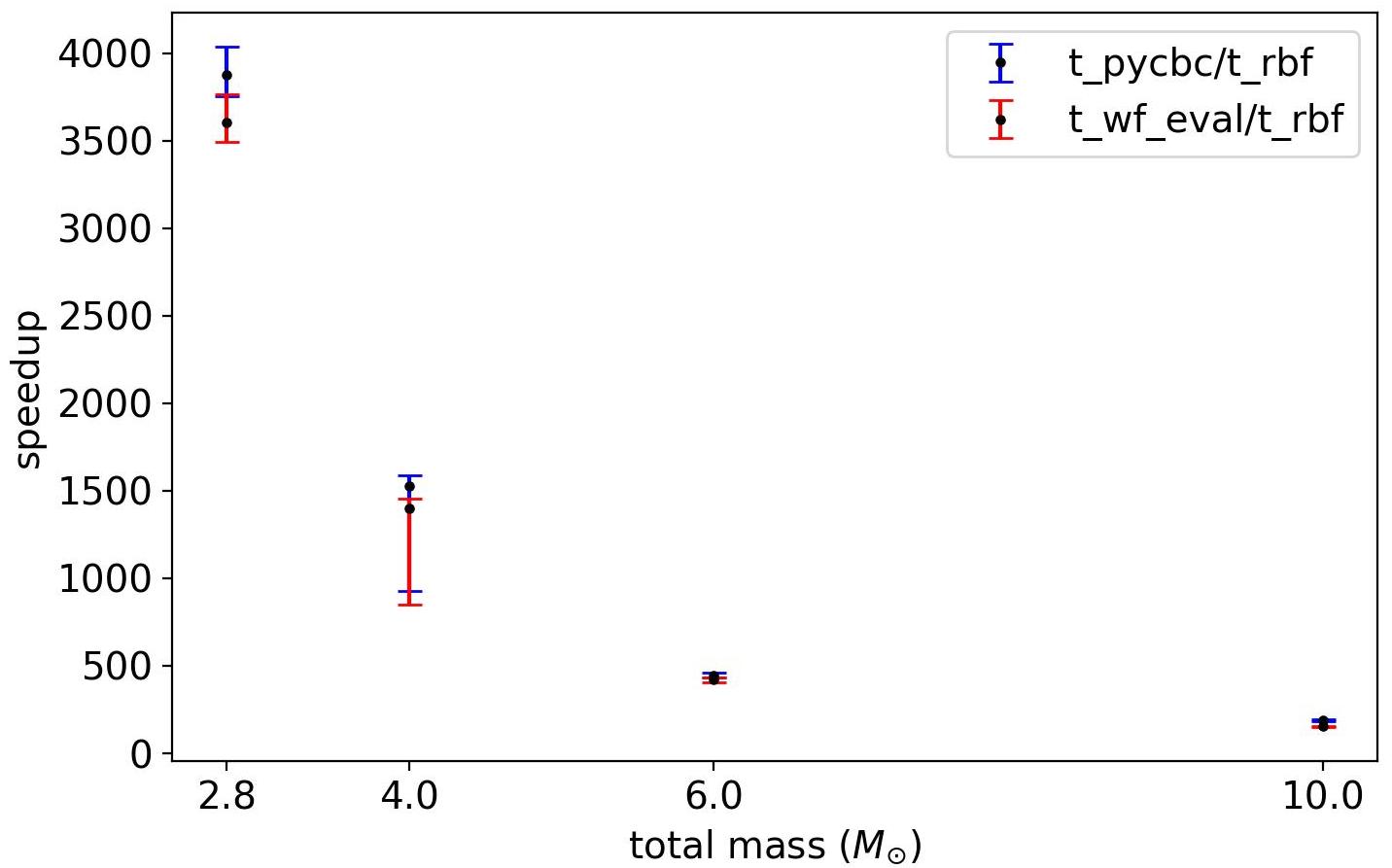}
\caption{The relative speed-up between the likelihood evaluation using PyCBC and the proposed meshfree method is shown (blue). Also, The ratio of overall time for generating waveforms (total mass: $2.8 M_{\odot} - 10 M_{\odot}$) using PyCBC and the likelihood evaluation via meshfree scheme has been compared (red). The dominant cost in the traditional likelihood calculation arises from the generation of long-duration waveforms.}
\label{Fig:speed-up-waveform}
\end{figure}

To verify this, we calculated the likelihood evaluation time using both PyCBC ($\text{t}_{\text{pycbc}}$) and meshfree method ($
\text{t}_{\text{rbf}}$) and also evaluated the waveform generation time (${\text{t}_{\text{wf}}}_{\text{eval}}$) for the four compact binary sources at $10^{4}$ points each. As shown in the plot below (Fig.~\ref{Fig:speed-up-waveform}), the waveform generation part is the dominant cost in traditional calculations. For a BNS system, it takes about $\sim 7$ times longer to generate the waveform as compared to the time for evaluating the likelihood integral. 
As we go towards heavy CBC systems, the speed-up ratio decreases as expected.
\end{document}